\documentclass[conference]{IEEEtran}
\ifCLASSINFOpdf
\else
\fi

\usepackage{subfigure}
\usepackage{caption}
\usepackage{graphics} 
\usepackage{epsfig} 
\usepackage{cite}
\usepackage{multirow}
\begin{document}
\title{Automatic Safety Helmet Wearing Detection}


\author{\IEEEauthorblockN{Kang Li, Xiaoguang Zhao, Jiang Bian, and Min Tan}
\IEEEauthorblockA{The State Key Laboratory of Management and Control for Complex System,\\ Institute of Automation, Chinese Academy of Sciences,\\ University of Chinese Academy of Sciences, Beijing, China\\ Email: {likang2014@ia.ac.cn, xiaoguang.zhao@ia.ac.cn, bianjiang2015@ia.ac.cn, min.tan@ia.ac.cn}}}
\maketitle

\begin{abstract}
Surveillance is very essential for the safety of power substation. The detection of whether wearing safety helmets or not for perambulatory workers is the key component of overall intelligent surveillance system in power substation. In this paper, a novel and practical safety helmet detection framework based on computer vision, machine learning and image processing is proposed. In order to ascertain motion objects in power substation, the ViBe background modelling algorithm is employed. Moreover, based on the result of motion objects segmentation, real-time human classification framework C4 is applied to locate pedestrian in power substation accurately and quickly. Finally, according to the result of pedestrian detection, the safety helmet wearing detection is implemented using the head location, the color space transformation and the color feature discrimination. Extensive compelling experimental results in power substation illustrate the efficiency and effectiveness of the proposed framework.
\end{abstract}


%
\IEEEpeerreviewmaketitle

\section{Introduction}

As is well known, a surveillance system is considerable significant for power substation safety. Over the past decades, some artificial intelligent techniques like computer vision and machine learning with growing progress has been widely applied in intelligent surveillance in power substation \cite{bib1}. It can not only avoid time consuming labour intensive task, but also point out the power equipment fault and worker illegal operation in time and accurately against accidents.

Wang \emph{et al.} designed a method to identify the status of the isolation switch and breaker in substation, which employed SIFT feature matching, Hough transform and KNN algorithm \cite{bib2}. Reddy \emph{et al.} proposed a scheme for condition monitoring of insulators. Based on the ROI region obtained from K-Means algorithm, this scheme used discrete orthogonal S-transform in conjunction with adaptive neuro-fuzzy inference system to ascertain the condition of the insulators \cite{bib3}. Chen \emph{et al.} presented a effective automatic detection and state recognition method for disconnecting switch, who makes full use of some prior knowledge about disconnecting switch and combines two important features of the fixed-contact \cite{bib4}. Liu \emph{et al.} developed an image-based state recognition approach: they propose extraction of texture features using a Gabor transformation, then the state of the isolator is classified by the SVM \cite{bib5}.

Above-mentioned researches mainly focus on the power equipment fault detection and state recognition. Aside from equipment safety, intelligent surveillance system still need to monitor operator safety work. The real time safety helmet wearing detection for perambulatory workers, as a most common safe operation situation in power substation, is a considerable important task related to worker safe. Thus it is necessary to develop a system for automatic detection of safety helmet wearing in power substation. Unfortunately, the related work is little and mostly had been made in the detection of motorcyclists with or without helmet. Waranusast \emph{et al.} utilized moving objects extracting and K-Nearest-Neighbor (KNN) classifier to develop a system which can automatically detect motorcycle riders and determine that they are wearing safety helmets or not \cite{bib6}. In \cite{bib7}, Silva \emph{et al.} applied the circular Hough transform and the Histogram of Oriented Gradients descriptor to extract feature, utilized the Multi-layer perceptron classifier to ascertain motorcyclist without helmet. There are little researches about detection of safety helmet wearing in power substation. In \cite{bib8}, the Kalman filtering and Cam-shift algorithm are used to track pedestrians and determine motion objects. Meanwhile, the color information of safety helmets is used to detect safety helmets wearing.

\begin{figure*}[!t]
    \centering
    \includegraphics[width=1\textwidth]{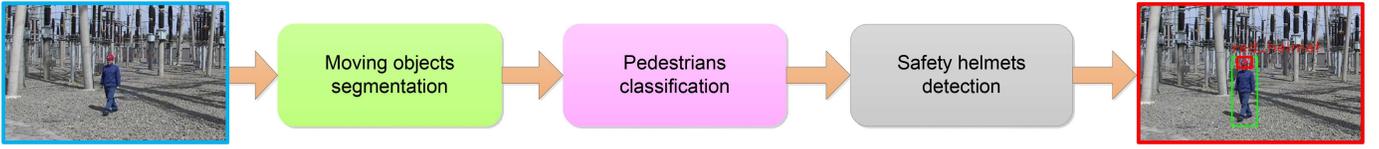}
    \caption{Safety helmet wearing detection system framework.}
    \label{fig:System}
\end{figure*}

The main purpose of this paper is to develop an innovative and practical safety helmet wearing detection system for perambulatory workers in power substation, based on computer vision, machine learning and image processing. The proposed detection framework focuses on design and implementation of detection of perambulatory workers with or without safety helmet in power substation, simultaneously is capable of ascertaining what color of the safety helmet. Toward this objective, the ViBe background modelling is employed to track motion objects. Furthermore, The C4 pedestrian detection algorithm is applied to determine motion objects whether human or not. Finally, the specific contribution made here is that utilizing the head location, the color space transformation and the color feature discrimination to detect whether wearing safety helmets or not for workers in power substation. Noticeably, the proposed framework is not only able to accurately realize the detection of whether wearing safety helmets or not for workers theoretically, but is also well appropriate of the practical application of intelligent surveillance in power substation on account of less dependence on hardware and more quick running speed.

The rest of this paper is organized as follows. Section II provides a overall framework of safety helmet wearing detection system. The background modeling, pedestrian detection and color feature discrimination method is detailed in Section III. Experiments and analyses are presented in Section IV. Finally, conclusions and future work are summarized in Section V.

\section{Safety Helmet Wearing Detection System Framework}

Fig.~\ref{fig:System} shows the schematic of safety helmet wearing detection system investigated in this paper. The system is divided in three steps: a) Moving objects segmentation; b) Pedestrians classification; c) Safety helmets detection.

\subsection{Moving Objects Segmentation}

The purpose of moving objects segmentation in power substation is that obtaining moving pixels as quick as possible to reduce time of searching region of interest. On account of the video camera fixed in power substation, the background information can be used as valuable prior knowledge to extract moving targets. The intuitive idea is that comparing per frame with background image, and then calculating the different distance in pixels to determination moving objects. Here, we using a fast background modeling algorithm named visual background extractor (ViBe)\cite{bib9}.

\subsection{Pedestrians classification}

Based on the results of moving objects segmentation, extracting image features and classifying pedestrians in power substation will become the most crucial task for safety helmet wearing detection. The more precisely pedestrians is classified, the accurately safety helmet is detected. Thus it is necessary to select an excellent feature and an efficient classifier. The strategy explored in this part is that using C4 pedestrian detection algorithm \cite{bib10}. The only difference of C4 algorithm in this paper is that C4 pedestrians classification is executed on the surrounding of motion objects rather than all pixels of raw frame.

\subsection{Safety helmets detection}

The final step is safety helmet detection. After the pedestrian classification, the head region need to be determined. Meanwhile, the color space transformation and color feature discrimination are implemented to ascertain whether wearing safety helmets or not for perambulatory workers. The image processing techniques including HSV transformation \cite{bib11} and adaptive threshold selection method \cite{bib12} are employed in this paper.

\section{Method Principles of System Framework}

\subsection{ViBe:a universal background subtraction algorithm}

ViBe is a pixel model used to estimate the background. The implementation of this algorithm is organized as three procedures: 1) model initialization 2) foreground segmentation 3) model update.

\subsubsection{Model Initialization}

Model initialization, as the first step of ViBe algorithm, uses only the first frame of video sequences. It is an attractive merit that differs from other background model initialization depending on known certain frame numbers of the video. The details will be described as follows.

Suppose that ${N_t}(x)$ is a set of $N$ neighborhood sample values at time $t$ of the pixel $x$, ${n_{ti}}(x)$ is the component of this sample set. The first frame sample set can be represented by:
\begin{equation}
\label{eq1}
{N_0}(x) = \{ {n_{01}}(x),{n_{02}}(x),...,{n_{0N}}(x)\}, i = 1,2,...,N
\end{equation}

For each pixel of the first frame, select randomly one value from ${N_0}(x)$ as the initial value of pixel $x$. This model initialization can be denoted as:
\begin{equation}
\label{eq2}
{B_0}(x) = \{ v|v \in {N_0}(x)\}
\end{equation}
where ${B_0}(x)$ is the initial background model of the pixel $x$, $v$ represents value of pixel.

\subsubsection{Foreground Segmentation}

Assume that ${S_R}({n_{ti}}(x))$ is a sphere of radius $R$ centered on ${n_{ti}}(x)$, which can be computed by comparing the ${n_{ti}}(x)$ to its closets values among the sample set ${N_t}(x)$. And then calculating the cardinality, defined $\#$, of the set intersection of this sphere and the sample set ${N_t}(x)$, which can be written as:
\begin{equation}
\label{eq3}
\# \{ {S_R}({n_{ti}}(x)) \cap {N_t}(x)\}
\end{equation}
Further, denoting $\# \min$ as a fixed threshold for background modeling. The pixel value is classified as foreground if the $\# \min$ is under the cardinality.

\subsubsection{Model Updating}

After the foreground detection, ViBe algorithm adopts a conservative update scheme. Firstly, background pixel has a possibility of ${1 \mathord{\left/{\vphantom {1 \beta }} \right. \kern-\nulldelimiterspace} \beta }$ to be updated. Moreover, there is only a certain ${1 \mathord{\left/{\vphantom {1 \beta }} \right. \kern-\nulldelimiterspace} \beta }$ possibility for the update of current frame instead of other method with updating per frame. Finally, allowing each pixel to diffuse between neighboring pixels, randomly select its own sample set for updating.

\subsection{C4:a real-time human detection framework}

\begin{figure}[!t]
    \centering
    \includegraphics[width=0.5\textwidth]{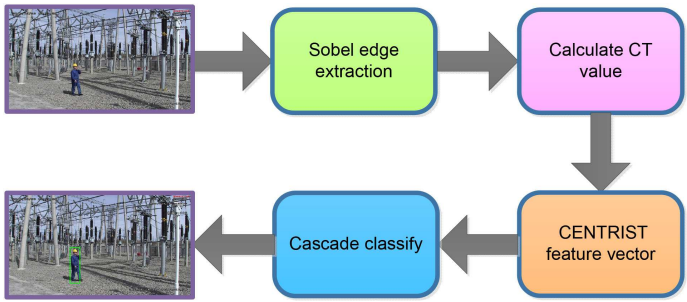}
    \caption{A overview of C4 human detection chain.}
    \label{fig:C4}
\end{figure}

C4 is a real-time and accurate human detector based on CENTRIST which focuses on contour cues for its detection. The overview of feature extraction and human detection chain is shown in Fig.~\ref{fig:C4}. This method shows that contour information is particularly crucial for pedestrian detection. Therefore, it uses Sobel edge detection \cite{bib13} and Census Transformation (CT) \cite{bib14} to encode contours of pedestrians. After that, C4 algorithm integrates the position information into feature vector by dividing the image patch into blocks and combining adjacent blocks into cells. At last, The cascade classifier of linear SVM and histogram intersection kernel (HIK) SVM \cite{bib15} is used to determine human or non-human. The more details are illustrated as follows.

\begin{figure}[!t]
    \centering
    \includegraphics[width=0.5\textwidth]{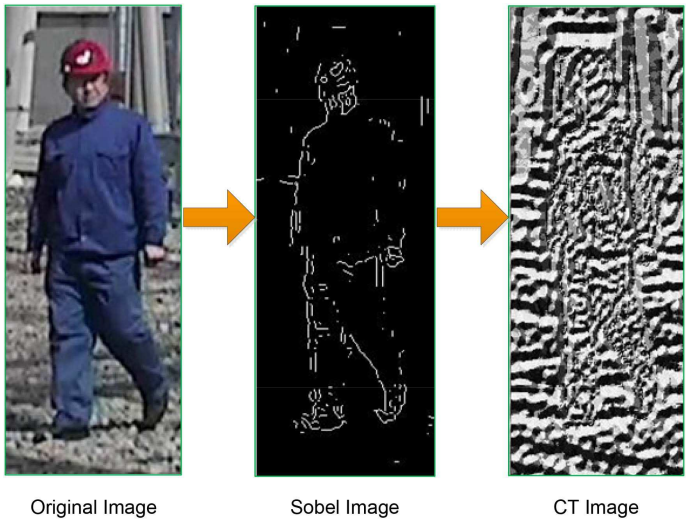}
    \caption{Human contour feature extraction. From left to right: \\original image, Sobel Image, CT image.}
    \label{fig:FeatureExtract}
\end{figure}

\subsubsection{Feature Extraction}

As is shown in Fig.~\ref{fig:FeatureExtract}, to find an excellent feature of human objects from the motion foreground, we compute the Sobel gradient of each pixel at first. And then an 8 bit value, named Census Transform, is calculated to encode the signs of comparison. As illustrated in Eq.~\ref{eq4}, by comparing a pixel value with its eight neighboring pixels, the bit is set to 1 if the center pixel is bigger than one of its neighbors or it is set to 0. The 8 bits can be collected in any order and converted to a CT value, which range from 0 to 255.

\begin{equation}
\label{eq4}
\begin{array}{*{20}{c}}
{32}&{64}&{96}\\
{32}&{64}&{96}\\
{32}&{32}&{96}
\end{array} \Rightarrow \begin{array}{*{20}{c}}
1&1&0\\
1&{}&0\\
1&1&0
\end{array} \Rightarrow {(11010110)_2} \Rightarrow CT = 214
\end{equation}

In order to introduce spatial distribution information for CENTRIST feature vector, C4 algorithm partitions the image patch into $9 \times 6$ blocks and regards any adjacent $2 \times 2$ blocks as a super-block. The extracting unit is a super-block, thus there are 24 super-blocks. The final feature vector is $256 \times 24 = 6144$ dimensions.

\subsubsection{Pedestrian Classification}

The detector is trained in a bootstrap way. In the training phase, we prepare a set of positive training image patches ${\cal P}$ and a set of non-human negative images ${\cal N}$. The first step is that randomly selecting a part of negative patches from ${\cal N}$ to form a subset of negative samples ${{\cal N}_1}$. Using ${\cal P}$ and ${{\cal N}_1}$, we train a SVM classifier ${H_1}$. Then, ${H_1}$ is employed to search hard examples, and these hard examples are added into ${{\cal N}_1}$ to form a new negative set ${H_2}$. Using ${\cal P}$ and ${{\cal N}_2}$, a SVM classifier ${H_2}$ is trained. It is a loop process until set times is over, the final trained linear SVM ${H_f}$ will be attained.

\begin{figure}[!t]
    \centering
    \includegraphics[width=0.4\textwidth]{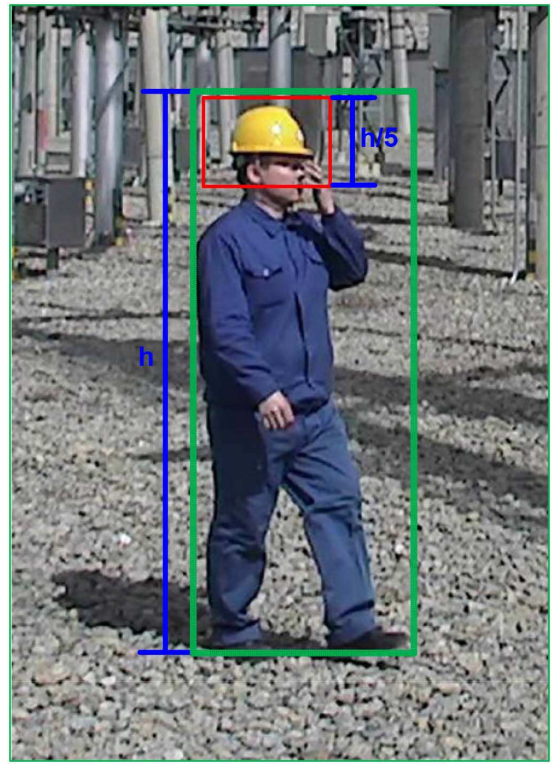}
    \caption{Interest of region location. The bounded red region is the ROI.}
    \label{fig:ROI}
\end{figure}

Here, the pedestrian classifier is a cascade classifier consisting of a linear SVM and a HIK SVM. Linear classifier, as the first classifier, achieves a rough determination of human with quick testing speed. While the HIK SVM classifier plays a more important role in accurate detection human on CENTRIST.

\subsection{Color feature discrimination}

Assume that the pedestrian classification is well fulfilled using C4, the safety helmet detection can make full use of the location information of pedestrian. As shown in Fig.~\ref{fig:ROI}, the top region of one-fifth in the bounding box is our interest of region for head location. This value is chosen empirically. In order to use the most important color information of helmet to determine human whether wearing helmet or not, color space transformation and color feature discrimination are executed. Since the image in HSV color space is more adaptive to color segmentation, so we convert RGB to HSV. It is worth noting that setting fix threshold for Hue and Saturation channels can segment various colors, and then discriminating human with or without safety helmet. Whereas we only set the Hue channel threshold and not define the Saturation channel threshold. Instead, we use OSTU method on the channel of Saturation to obtain automatically threshold and segment color.

\section{Results And Discussions}

\subsection{Experimental Setup}

To evaluate the proposed safety helmet detection framework, extensive experiments were conducted on surveillance video sequences of power substation, which contain single pedestrian and multiple pedestrians under challenging backgrounds including electrical wire, wire pole, railing and etc. We capture ten videos with different scenes under fixed view and run our detection algorithm. Noticeably, proposed method is only adapted to detect whether wearing safety helmet or not for perambulatory workers, because safety patrol is one of the most common situations in power substation. Therefore, ten videos are both perambulatory situations.

Here, we have labeled data into wearing safety helmet and no wearing safety helmet manually. The accuracy of pedestrians classification is defined as $Ac{c_{pd}} = \frac{T}{{T + F}}$, where $T$ represents the number of correct classification results for pedestrians, $F$ represents the number of false classification results for pedestrians. Besides, based on the accurate pedestrian classification, Receiver Operating Characteristic (ROC) curve and Precision Recall (PR) curve of safety helmet wearing detection system are drew to illustrate the excellent performance of proposed method.

\subsection{Experimental Results and Analyses}

\begin{figure}[!htp]
    \centering
    \subfigure[Motion objects detection on scenes including single pedestrian]{
    \begin{minipage}[b]{0.4\textwidth}
    \includegraphics[width=1\textwidth]{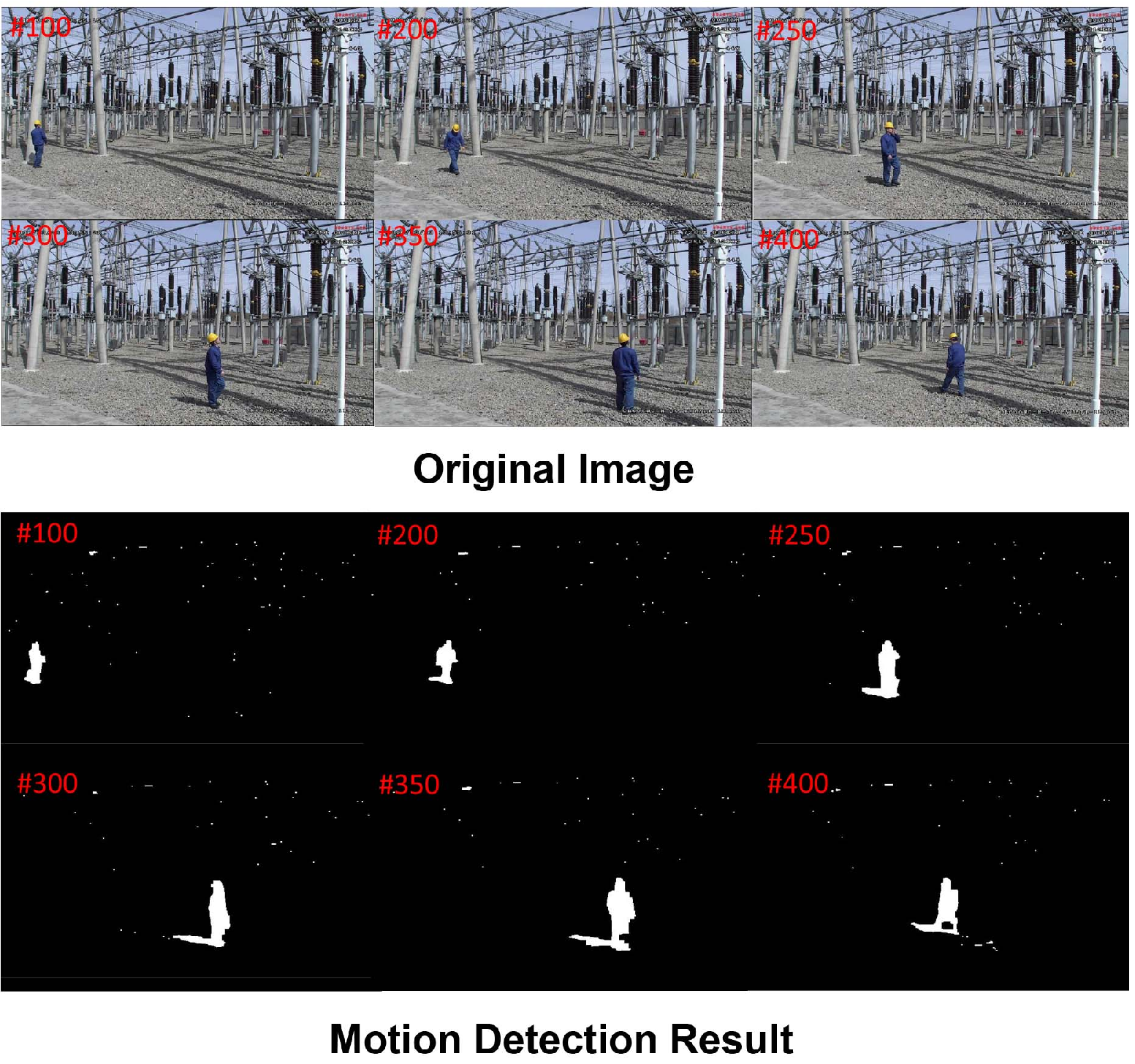}
    \end{minipage}
    }
    \subfigure[Motion objects detection on scenes including several pedestrians]{
    \begin{minipage}[b]{0.4\textwidth}
    \includegraphics[width=1\textwidth]{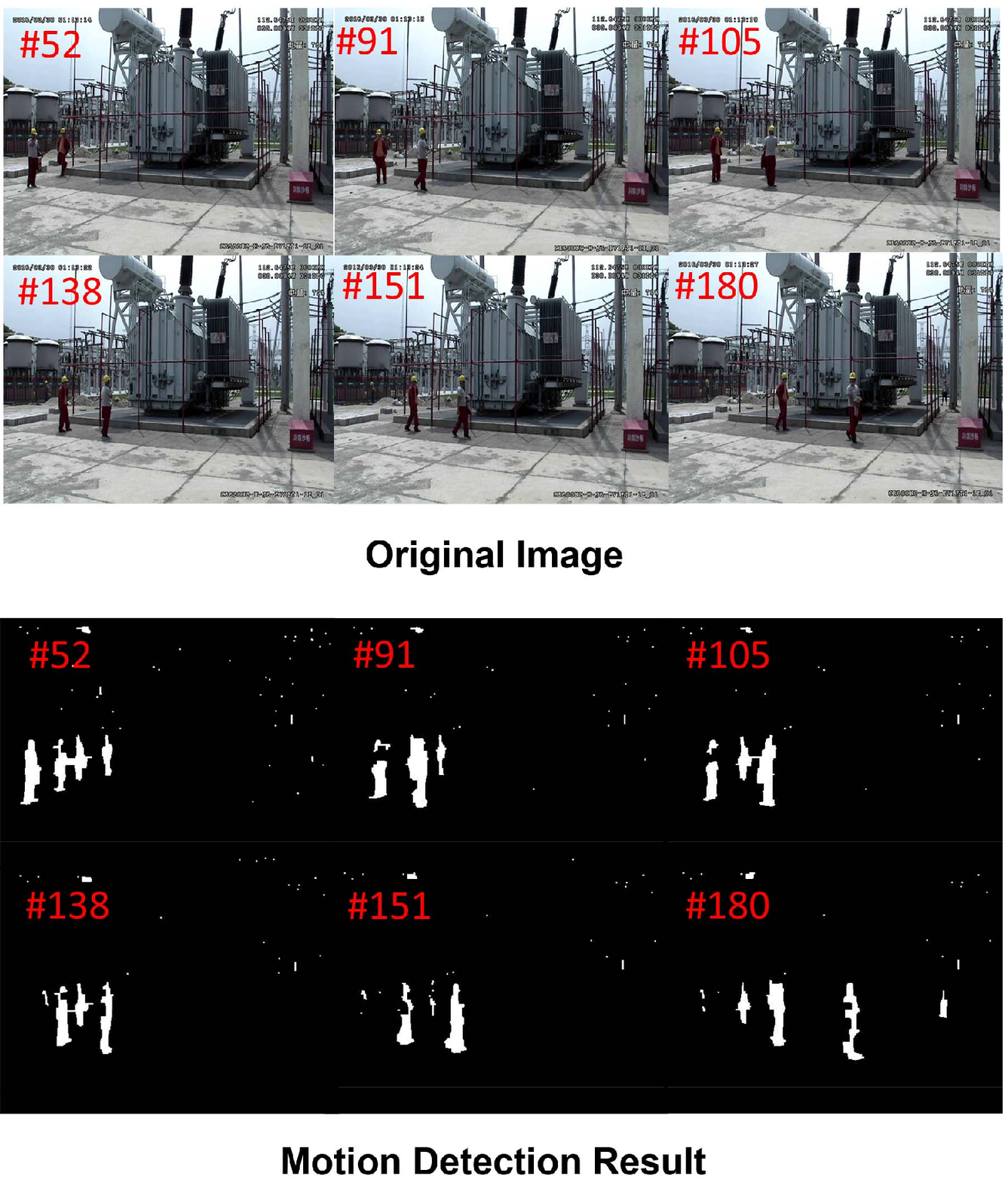}
    \end{minipage}
    }
    \caption{Motion objects detection results.}
    \label{fig:Background}
\end{figure}

\begin{figure}[!htp]
    \centering
    \subfigure[Single pedestrian classification]{
    \begin{minipage}[b]{0.4\textwidth}
    \includegraphics[width=1\textwidth]{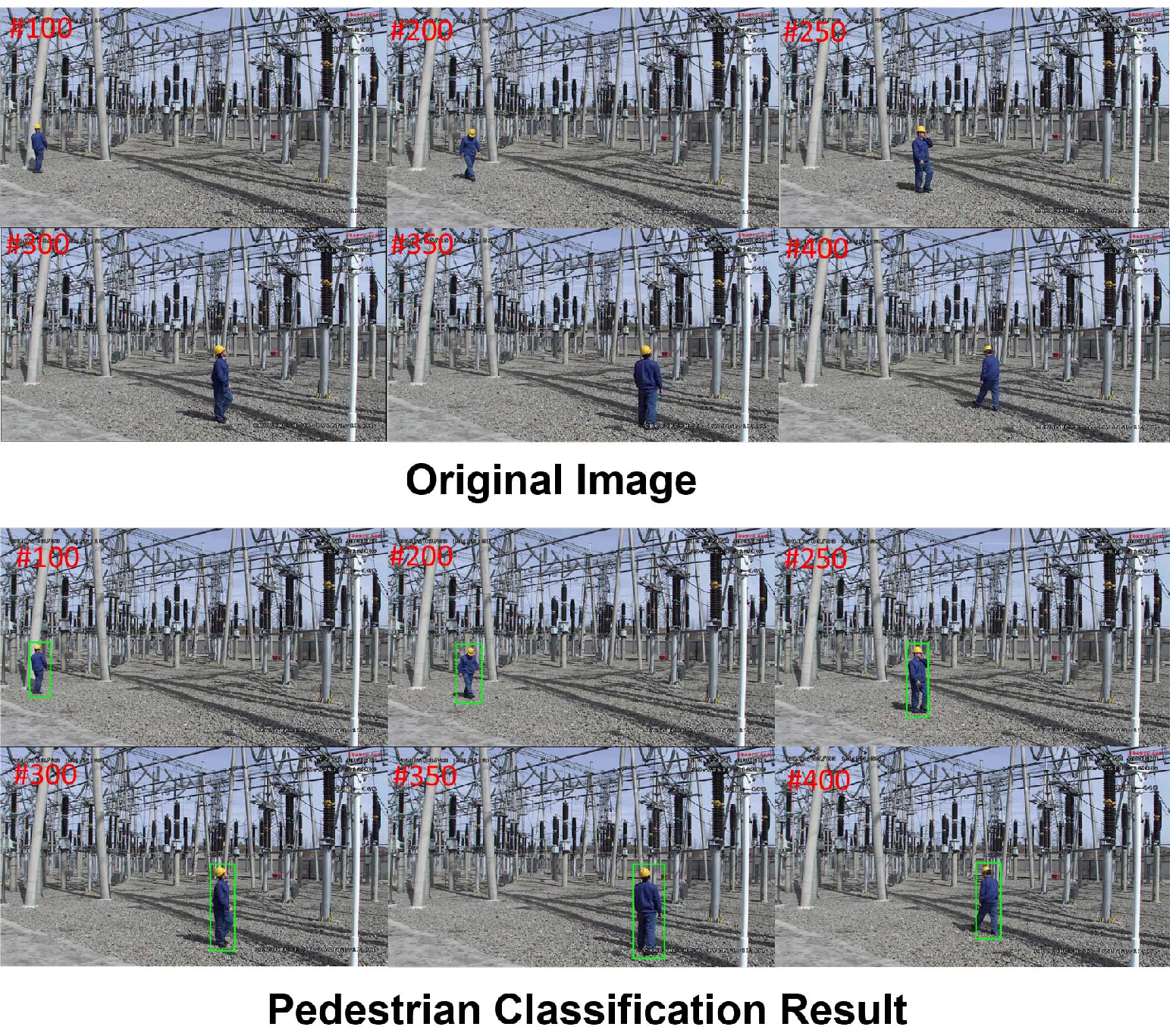}
    \end{minipage}
    }
    \subfigure[Multiple pedestrians classification]{
    \begin{minipage}[b]{0.4\textwidth}
    \includegraphics[width=1\textwidth]{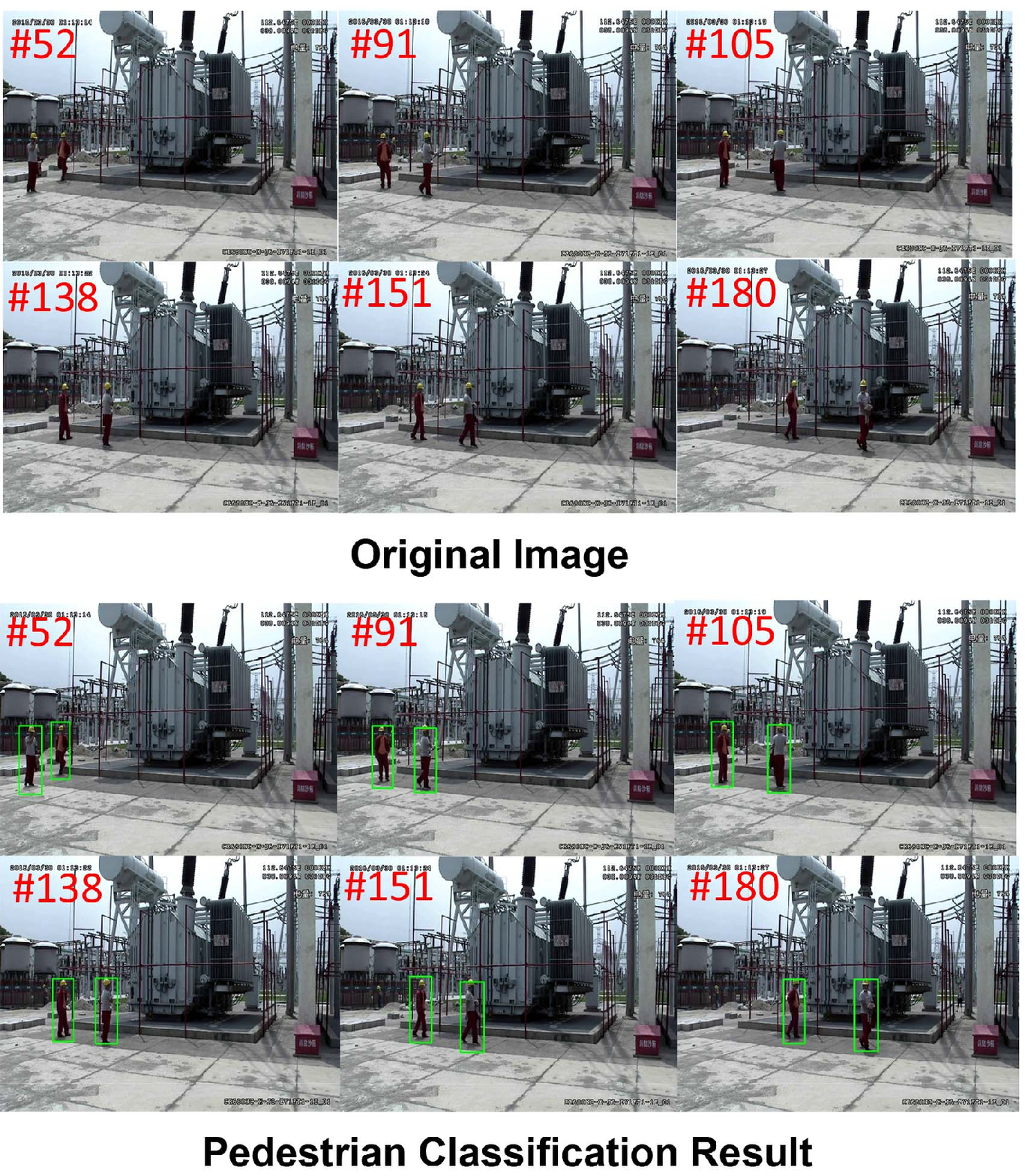}
    \end{minipage}
    }
    \caption{Pedestrian detection results.}
    \label{fig:Pedestrians}
\end{figure}

\begin{figure}[!htp]
    \centering
    \subfigure[Single pedestrian safety helmet detection]{
    \begin{minipage}[b]{0.4\textwidth}
    \includegraphics[width=1\textwidth]{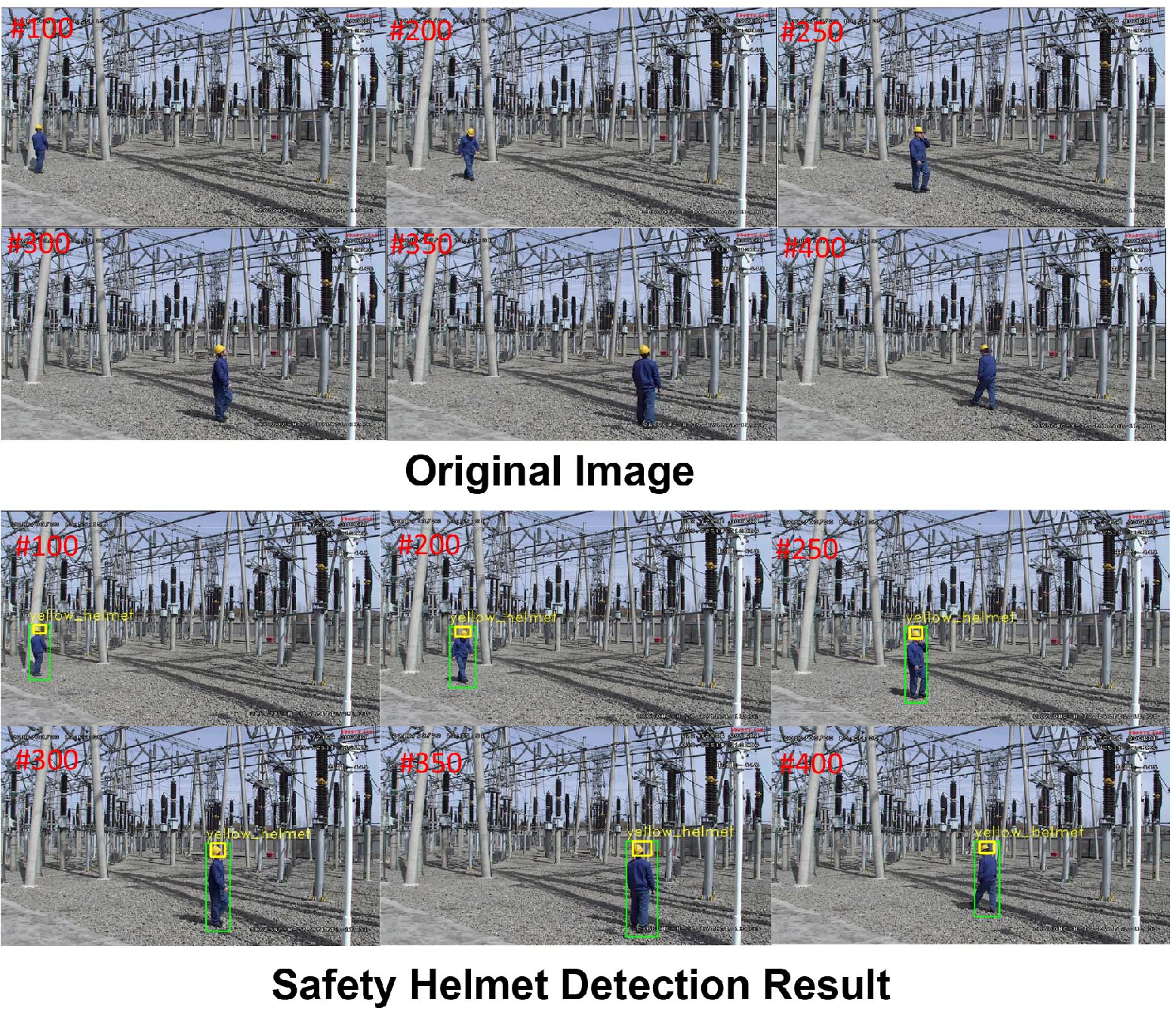}
    \end{minipage}
    }
    \subfigure[Multiple pedestrians safety helmets detection]{
    \begin{minipage}[b]{0.4\textwidth}
    \includegraphics[width=1\textwidth]{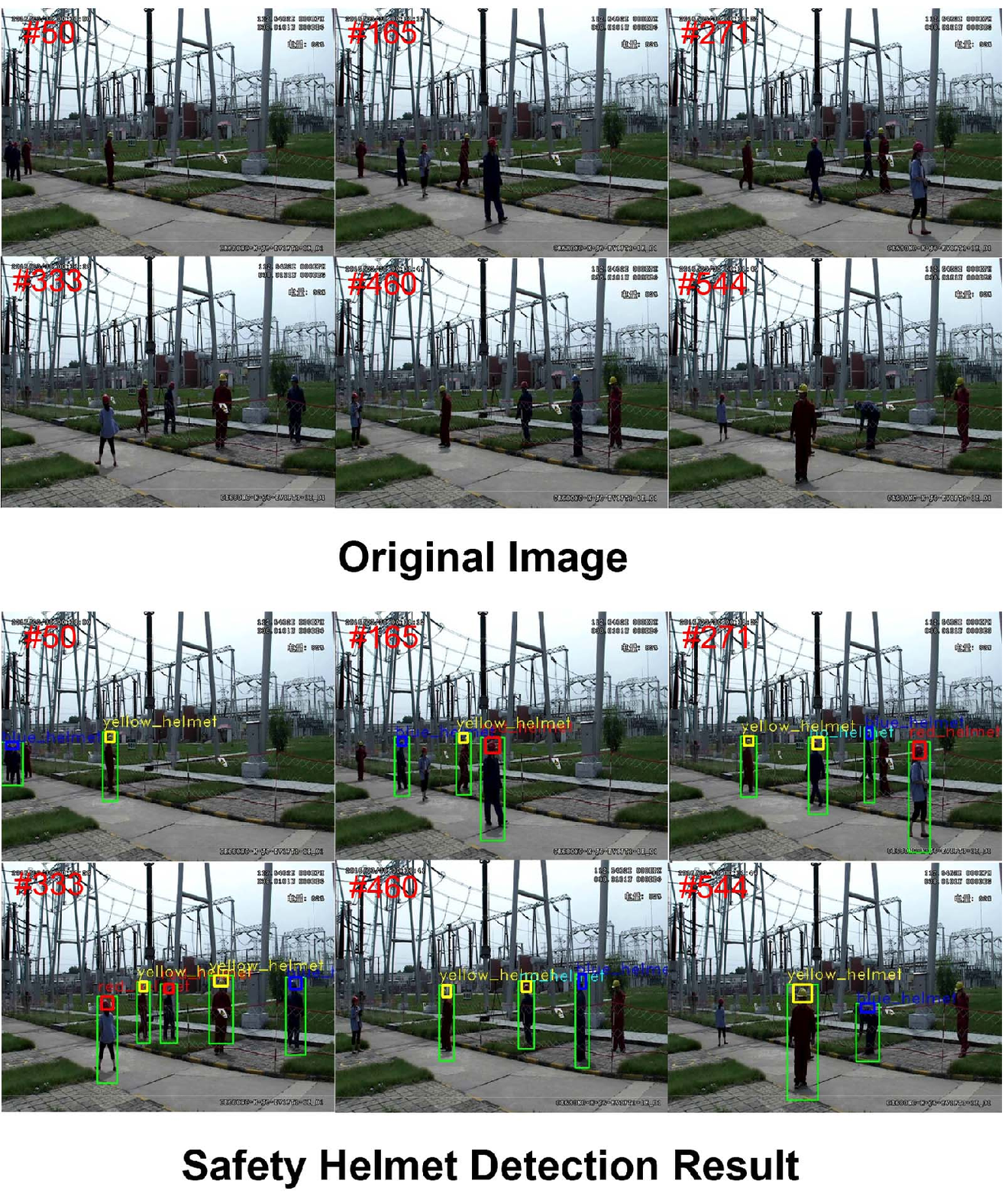}
    \end{minipage}
    }
    \caption{Safety helmet detection results.}
    \label{fig:Helmets}
\end{figure}

Background modelling is the first step of safety helmet wearing detection. As is shown in Fig.~\ref{fig:Background} $(a)$ and $(b)$, the motion objects on scenes consisting of single pedestrian and multiple pedestrians respectively are segmented. The white pixels are motion foregrounds, the black pixels are static background instead. The locomotor objects detection results were satisfactory.

In terms of pedestrian detection, we use the \emph{INRIA} person dataset \cite{bib16} and pedestrian data collected from power substation to train classifier. Fig.~\ref{fig:Pedestrians} $(a)$ and $(b)$ show the classification results of pedestrian on scenes including single pedestrian and multiple pedestrians respectively. The performance of the classifier is excellent on power substation with complex background.

Finally, the results of safety helmet wearing detection were presented in Fig.~\ref{fig:Helmets} $(a)$ and $(b)$. Single pedestrian safety helmet can be detected pretty well, but multiple pedestrians safety helmets maybe not detected sometimes. In our opinion, the main reason of failed examination is that the target is far from the surveillance camera. C4 pedestrian classification can not work on the too little target. Meanwhile, C4 features have limit description capability for pedestrians, which also can lead to failed examination. Besides, there are two reasons of false detection. One is the errors of pedestrian detection. such as the too big or too small bounding box, which can cause the error location of head. The other is that the color feature of safety helmet is sensitive to illumination changes. So here are much improving space.

\begin{figure}[!t]
    \centering
    \includegraphics[width=0.4\textwidth]{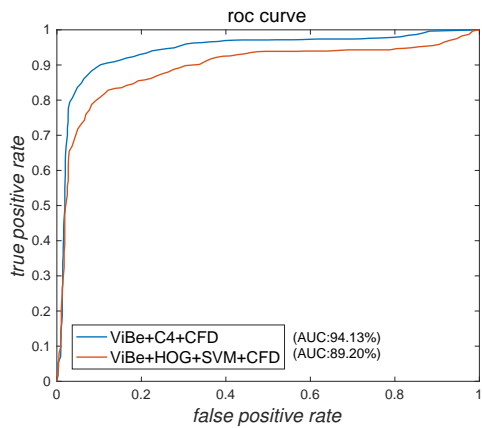}
    \caption{ROC curve for all classifiers in helmet detection.}
    \label{fig:ROC}
\end{figure}

\begin{figure}[!t]
    \centering
    \includegraphics[width=0.4\textwidth]{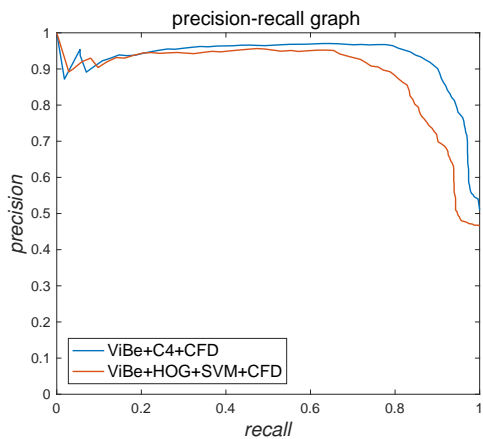}
    \caption{PR curve for all classifiers in helmet detection.}
    \label{fig:PR}
\end{figure}

Aside from qualitative analysis, quantitative analysis is needful to evaluate the performance of proposed method. The mean accuracy of pedestrian classification $Ac{c_{pd}}$ in ten videos is around 84.2$\%$. Fig.~\ref{fig:ROC} and Fig.~\ref{fig:PR} respectively show the ROC curve and PR curve of different methods on ten videos in power substation. We can see that the performance of ViBe algorithm in conjunction with C4 classifier and color feature discrimination (CFD) is better than ViBe algorithm combining with HOG feature extraction, support vector machine (SVM) and color feature discrimination (CFD). The AUC index of ROC curve is used to determine the quality of different classifiers. The AUC of method using HOG feature extraction and SVM classifier is 89.20$\%$, and the AUC of our method is 94.13$\%$. It improves around 5$\%$.

In a word, no matter qualitative analysis or quantitative analysis, the proposed method has achieved accurate detection of perambulatory workers whether wearing safety helmet or not to a certain extent.

\subsection{Discussion}

This work is meaningful for safety surveillance because the patrol and examine is most common in power substation and the determination of wearing safety helmet is crucial for life safe of perambulatory workers. In order to liberate labors, computer vision and machine learning algorithm is employed to complete this detection task. The ViBe background algorithm shrinks searching range for perambulatory workers. C4 algorithm implements pedestrians classification quickly. Color feature discrimination (CFD) ascertain safety helmet wearing situations. The overall framework is efficient and effective for this task.

However, there are some challenging problems. For example, Method of C4 pedestrian classification in conjunction with color feature discrimination with fixed parameters does not work very well for varying environments such as weather changes, which needs to modify parameters in different scenes.

\section{Conclusion}

In this paper we have developed a novel and practical safety helmet wearing detection system to determine perambulatory workers to whether wear safety helmet or not. ViBe background modelling algorithm has realized excellent motion object detection. Furthermore, C4 pedestrians classification algorithm has implemented the worker location quickly. The crucial step of proposed safety helmet detection framework is color feature discrimination on HSV color space, which has achieved a great performance on common situations in power substation. Extensive experimental results has illustrated the effectiveness and efficiency of this safety helmet wearing detection system.

The ongoing and future work will focus on safety helmet wearing detection on some scenes including multiple pedestrians and complex background. The more accurate feature design and more robust detection method explore will be executed in future.


\section*{Acknowledgment}

This work was partly supported by National Natural Science Foundation of China under Grants 61271432 and 61421004, and partly supported by the State Grid Corporation of China Technology Projects(52053015000G).



%

\end{document}